\documentclass[prb,twocolumn,showpacs]{revtex4}
\usepackage{graphicx} 

\usepackage{amsmath,amsthm,amssymb,mathrsfs}
\usepackage{bm}

\begin{document}

\title{Dissipation due to pure spin-current generated by spin pumping}

\author{Tomohiro Taniguchi$^{1,3}$ and Wayne M. Saslow$^{2,3}$
      }
 \affiliation{
 $^{1}$ 
 National Institute of Advanced Industrial Science and Technology (AIST), Spintronics Research Center, Tsukuba, Ibaraki 305-8568, Japan, 
 \\
 $^{2}$ Department of Physics, Texas A\&M University, College Station, Texas, 77843-4242, U.S.A., 
 \\
 $^{3}$ Center for Nanoscale Science and Technology, National Institute of Standards and Technology, Gaithersburg, Maryland, 20899-6202, U.S.A.
 }

 \date{\today} 
 \begin{abstract}
  {
    Based on spin-dependent transport theory and thermodynamics, 
    we develop a generalized theory of the Joule heating 
    in the presence of a spin current. 
    Along with the conventional Joule heating consisting of an electric current and electrochemical potential, 
    it is found that the spin current and spin accumulation give an additional dissipation 
    because the spin-dependent scatterings inside bulk and ferromagnetic/nonmagnetic interface lead to 
    a change of entropy. 
    The theory is applied to investigate 
    the dissipation due to pure spin-current generated by spin pumping 
    across a ferromagnetic/nonmagnetic/ferromagnetic multilayer. 
    The dissipation arises from an interface 
    because the spin pumping is a transfer of both 
    the spin angular momentum and the energy from the ferromagnet to 
    conduction electrons near the interface. 
    It is found that the dissipation is proportional 
    to the enhancement of the Gilbert damping constant by spin pumping. 
  }
 \end{abstract}

 \maketitle


\section{Introduction}
\label{sec:Introduction}

Dissipation due to electron transport in a conductor is an important issue 
for both fundamental and applied physics [\onlinecite{johnson87,parrott96,saslow07,sears11,tulapurkar11}].
According to electron transport theory [\onlinecite{rammer08}], 
the conductivity of the electron becomes finite 
because of impurity scattering inside the conductor, 
which leads to Joule heating $J_{e}E$,
where $J_{e}$ and $E$ are the electric current density and electric field, respectively. 
Motivated to reduce power consumption due to Joule heating, 
as well as because of a fundamental interest in its quantum mechanical nature, 
the generation of a pure spin-current by spin pumping, spin-Seebeck effect, or spin-Hall effect  
has been extensively investigated [\onlinecite{mizukami02a,mizukami02b,ando08,ando12,uchida08,liu12a,liu12b,miao13,ando13}]. 


Dissipation is associated with the production of entropy. 
Spin-flip processes and spin-dependent scatterings within a bulk ferromagnet (F) or nonmagnet (N) 
and at an F/N interface mix the spin-up and spin-down states, 
leading to a change of the entropy. 
Therefore a physical system, 
such as a F/N metallic multilayer, carrying a pure spin-current, 
still dissipates energy 
even in the absence of an electric current. 
A quantitative evaluation of the dissipation 
due to pure spin-current therefore is a fundamentally important problem. 


In 1987, Johnson and Silsbee [\onlinecite{johnson87}] studied 
the surface and bulk transport coefficients for spin conduction, 
and the associated entropy production rates, 
without considering the rate of interface heating. 
More recently, Sears and Saslow [\onlinecite{sears11}] used irreversible thermodynamics 
to study interface heating due to electric current in a magnetic system, 
and Tulapurkar and Suzuki [\onlinecite{tulapurkar11}] used the Boltzmann equation 
to investigate bulk and interface heating for spin conduction. 
Reference [\onlinecite{tulapurkar11}] shows that, 
roughly speaking, the dissipation due to spin current is 
proportional to the square of the spin polarization of the conduction electrons, 
indicating that the heating associated with the spin current is 
much smaller than that due to the electric current. 
However, these works consider only a collinear alignment of 
the magnetizations in a F/N multilayer, 
so only the longitudinal components of the spin current and spin accumulation 
(i.e., spin chemical potential, proportional to the nonequilibrium spin density) appear. 
({\it Longitudinal} and {\it transverse} will be used to mean that the direction of the spin polarization is 
collinear or normal to the local magnetization.) 
On the other hand, in many physical phenomena, 
such as spin torque switching [\onlinecite{slonczewski96}] and spin pumping [\onlinecite{mizukami02a,mizukami02b}], 
a non-collinear alignment of the magnetizations generally appears, 
in which transverse spin current and spin accumulation exist. 
For example, spin pumping is a generation of the transverse spin current 
by the transfer of spin angular momentum from the ferromagnetic layer to the conduction electrons 
[\onlinecite{mizukami02a,mizukami02b,silsbee79,tserkovnyak02a,tserkovnyak02b,tserkovnyak03,tserkovnyak05,takahashi14}]. 
Bulk heating due to spin pumping 
in a magnetic wire within a domain wall (driven by $\mathbf{m}\times\mathbf{H}$) has also been studied [\onlinecite{saslow07}], 
but was not extended to include interface heating. 
In these works, the main contribution to the dissipation arises from the electric current. 
The present work develops a unified theory of dissipation 
which enables the simultaneous evaluation of both bulk and interface heating 
in a ferromagnetic system, with the spin current 
having arbitrary alignment of the magnetizations. 
Also, an evaluation of the dissipation due to a pure spin-current is indispensable for comparison 
with experiments that determine the rate of heating. 


This paper develops a general theory of dissipation 
in the presence of spin current 
based on the spin-dependent transport theory and thermodynamics. 
It is found that, along with the conventional Joule heating, 
the spin current $\mathbf{I}_{s}$ (or its density $\mathbf{J}_{s}$) 
and spin accumulation $\bm{\mu}$ contribute to the bulk and interface dissipations, 
as shown in Eqs. (\ref{eq:dissipation_volume}) and (\ref{eq:dissipation_area}). 
We apply the theory to evaluate the dissipation 
due to a pure spin-current generated by spin pumping 
in the ferromagnetic (F${}_{1}$) / nonmagnetic (N) / ferromagnetic (F${}_{2}$) multilayer. 
Spin pumping provides an interesting example to study 
the dissipation problem of pure spin-current. 
In spin pumping, electric current is absent throughout the system. 
The electron transport is described by a one-dimensional equation, 
and an external temperature gradient is absent, 
which makes evaluation of the dissipation simple 
compared with the spin-Seebeck effect or spin-Hall effect.
It is found that the dissipation is proportional to 
the enhancement of the Gilbert damping by spin pumping. 
The amount of the dissipation due to the spin pumping is maximized 
for an orthogonal alignment of the two magnetizations. 
For the conditions we study, the maximum dissipation is estimated to 
be two to three orders of magnitude smaller than the dissipation 
due to the electric current when there is spin torque switching. 


The paper is organized as follows. 
In Sec. \ref{sec:Spin pumping in F/N/F system}, 
the system we consider is illustrated. 
Section \ref{sec:Dissipation formulas} formulates a theory of dissipation 
of spin-polarized conduction electrons, using diffusive spin transport theory and thermodynamics. 
Section \ref{sec:Dissipation due to spin pumping} studies the relationship between the dissipation due to spin pumping 
and the equation developed in the previous section.  
Section \ref{sec:Evaluation of dissipation} quantitatively evaluates the dissipation due to spin pumping.
Section \ref{sec:Comparison with spin torque switching}, 
compares the spin pumping dissipation with the dissipation in the case of spin torque switching. 
Section \ref{sec:Conclusion} provides our conclusions.




\begin{figure}
\centerline{\includegraphics[width=0.8\columnwidth]{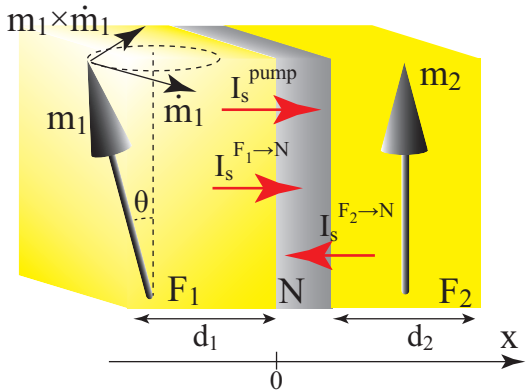}}\vspace{-3.0ex}
\caption{
         Schematic view of the F${}_{1}$/N/F${}_{2}$ ferromagnetic multilayer system. 
         The directions of $\dot{\mathbf{m}}_{1}$ and $\mathbf{m}_{1}\times\dot{\mathbf{m}}_{1}$ are indicated by arrows. 
         \vspace{-3ex}}
\label{fig:fig1}
\end{figure}





\section{Spin pumping in F/N/F system}
\label{sec:Spin pumping in F/N/F system}

Figure \ref{fig:fig1} shows a schematic view of the F${}_{1}$/N/F${}_{2}$ ferromagnetic multilayer system, 
where $\mathbf{m}_{1}$ and $\mathbf{m}_{2}$ are unit vectors 
pointing along the magnetizations of the F${}_{1}$ and F${}_{2}$ layers, respectively. 
Where needed, subscripts $k=1,2$ denote the F${}_{k}$ layer. 
The thickness of the F${}_{k}$ layer is denoted by $d_{k}$. 
The F${}_{1}$ and F${}_{2}$ layers lie in the regions 
$-d_{1} \le x \le 0$ and $0 \le x \le d_{2}$, respectively. 
We assume that the spin current is conserved in the N layer, 
and thus consider its zero-thickness limit
because a typical value for the spin diffusion length of an N layer 
is much greater than its thickness: 
for example, the spin diffusion length for Cu is on the order of 100 nm, 
whereas experimental thicknesses are less than 5 nm [\onlinecite{mizukami02a,mizukami02b,bass07}]. 


Steady precession of $\mathbf{m}_{1}$ with the cone angle $\theta$ can be excited 
by microwave radiation of the angular velocity $\omega$ 
for ferromagnetic resonance (FMR) in the F${}_{1}$ layer. 
Then, the F${}_{1}$ layer pumps the pure spin-current 
\begin{equation}
  \mathbf{I}_{s}^{\rm pump}
  =
  \frac{\hbar}{4\pi}
  \left(
    g_{\rm r(F_{1})}^{\uparrow\downarrow}
    \mathbf{m}_{1}
    \times
    \frac{d \mathbf{m}_{1}}{dt}
    +
    g_{\rm i(F_{1})}^{\uparrow\downarrow}
    \frac{d \mathbf{m}_{1}}{dt}
  \right), 
  \label{eq:spin_pumping}
\end{equation}
where 
the real and imaginary parts of the mixing conductance are denoted by 
$g_{\rm r}^{\uparrow\downarrow}$ and $g_{\rm i}^{\uparrow\downarrow}$, respectively [\onlinecite{brataas01,brataas06}]. 
The pumped spin current creates spin accumulations 
in the ferromagnetic ($\bm{\mu}_{\rm F}$) and nonmagnetic ($\bm{\mu}_{\rm N}$) layers, 
which induce backflow spin current (into N) [\onlinecite{tserkovnyak03,brataas01,brataas06,taniguchi08a}], 
given by 
\begin{equation}
\begin{split}
  \mathbf{I}_{s}^{\rm F \to N}
  =
  \frac{1}{4\pi}
  &
  \left[
    \frac{(1-\gamma^{2})g}{2}
    \mathbf{m}
    \cdot
    \left(
      \bm{\mu}_{\rm F}
      -
      \bm{\mu}_{\rm N}
    \right)
    \mathbf{m}
  \right.
\\
  & -
    g_{\rm r}^{\uparrow\downarrow}
    \mathbf{m}
    \times
    \left(
      \bm{\mu}_{\rm N}
      \times
      \mathbf{m}
    \right)
    -
    g_{\rm i}^{\uparrow\downarrow}
    \bm{\mu}_{\rm N}
    \times
    \mathbf{m}
\\
  &+
  \left.
    t_{\rm r}^{\uparrow\downarrow}
    \mathbf{m}
    \times
    \left(
      \bm{\mu}_{\rm F}
      \times
      \mathbf{m}
    \right)
    +
    t_{\rm i}^{\uparrow\downarrow}
    \bm{\mu}_{\rm F}
    \times
    \mathbf{m}
  \right].
  \label{eq:backflow} 
\end{split}
\end{equation}
The total interface conductance 
$g=g^{\uparrow\uparrow}+g^{\downarrow\downarrow}$ and 
the spin polarization of the interface conductance 
$\gamma=(g^{\uparrow\uparrow}-g^{\downarrow\downarrow})/(g^{\uparrow\uparrow}+g^{\downarrow\downarrow})$ are 
defined from the interface resistance of the spin-$\nu$ ($\nu=\uparrow,\downarrow$) electrons 
$r^{\nu\nu}=(h/e^{2})S/g^{\nu\nu}$, 
where $S$ is the cross section area. 
The real and imaginary parts of the transmission mixing conductance 
at the F/N interface are denoted by $t_{\rm r(i)}^{\uparrow\downarrow}$. 
The condition that the spin current is conserved in the N layer can be expressed as 
\begin{equation}
  \mathbf{I}_{s}^{\rm pump}
  +
  \mathbf{I}_{s}^{\rm F_{1} \to N} 
  + 
  \mathbf{I}_{s}^{\rm F_{2} \to N}
  =
  \bm{0}. 
  \label{eq:conservation_spin_current}
\end{equation}



\section{Dissipation formulas}
\label{sec:Dissipation formulas}

To obtain the dissipation due to spin pumping, 
it is necessary to investigate how the spin accumulation relaxes 
inside the F layers and at the F/N interfaces. 
For generality we include the terms related to the electric current and field, 
although these are absent in the spin-pumped system. 
The spin accumulation in the ferromagnetic layer relates to 
the distribution function $\hat{F}=(f_{0}+\mathbf{f}\cdot\bm{\sigma})/2$, 
which is a $2\times 2$ matrix in spin space 
and satisfies the Boltzmann equation [\onlinecite{tulapurkar11,taniguchi08a,valet93,simanek01,zhang02,shpiro03,zhang04,piechon07,taniguchi09}], 
via [\onlinecite{comment1}] $\bm{\mu}= \int_{\varepsilon_{\rm F}} {\rm Tr}[\bm{\sigma}\hat{F}] d \varepsilon$, 
$\bm{\sigma}$ being the Pauli matrices. 
The charge and spin distributions are 
denoted by $f_{0}$ and $\mathbf{f}$, respectively. 
The distributions for spin parallel, $f_{\uparrow}=(f_{0}+\mathbf{m}\cdot\mathbf{f})/2$, 
or antiparallel, $f_{\downarrow}=(f_{0}-\mathbf{m}\cdot\mathbf{f})/2$, to 
the local spin, give the longitudinal spin. 
On the other hand, the components of $\mathbf{f}$ 
orthogonal to $\mathbf{m}$ correspond to the transverse spin. 
Below, we introduce the following notations 
to distinguish the longitudinal ("L") and transverse ("T") components of 
the spin current $\mathbf{I}_{s}$ and spin accumulation $\bm{\mu}$: 
\begin{align}
&
  \mathbf{I}_{s}^{\rm L}
  =
  \left(
    \mathbf{m}
    \cdot
    \mathbf{I}_{s}
  \right)
  \mathbf{m},
  \label{eq:longitudinal_spin_current_def}
\\
&
  \mathbf{I}_{s}^{\rm T}
  =
  \mathbf{m}
  \times
  \left(
    \mathbf{I}_{s}
    \times
    \mathbf{m}
  \right),
  \label{eq:transverse_spin_current_def}
\\
&
  \bm{\mu}^{\rm L}
  =
  \left(
    \mathbf{m}
    \cdot
    \bm{\mu}
  \right)
  \mathbf{m},
  \label{eq:longitudinal_spin_accumulation_def}
\\
&
  \bm{\mu}^{\rm T}
  =
  \mathbf{m}
  \times
  \left(
    \bm{\mu}
    \times
    \mathbf{m}
  \right),
  \label{eq:transverse_spin_accumulation_def}
\end{align}
where $\mathbf{I}_{s}$ equals to 
$\mathbf{I}_{s}^{\rm pump}+\mathbf{I}_{s}^{\rm F_{1} \to N}$ at the F${}_{1}$/N interface 
and $-\mathbf{I}_{s}^{\rm F_{2} \to N}$ at the F${}_{2}$/N interface, respectively. 
The spin current density is denoted as $\mathbf{J}_{s}=\mathbf{I}_{s}/S$. 


We first consider the diffusive transport for the longitudinal spin [\onlinecite{valet93,simanek01,zhang02,shpiro03,zhang04,piechon07,taniguchi09}]. 
The longitudinal spin accumulation relates to the electrochemical potential 
$\bar{\mu}_{\nu}=\mu_{0}+\delta\mu_{\nu}-eV$ ($\nu=\uparrow,\downarrow$) via 
$\bm{\mu}^{\rm L}=(\bar{\mu}_{\uparrow}-\bar{\mu}_{\downarrow})\mathbf{m}$, 
where $\mu_{0}$, $\delta\mu_{\nu}$, and $-eV$ are 
the chemical potential in equilibrium, 
its deviation in nonequilibrium, 
and the electric potential. 
The longitudinal electron density $n_{\nu}=\int d^{3}\mathbf{k}/(2\pi)^{3} f_{\nu}$ 
and its current density $j_{\nu}=\int d^{3}\mathbf{k}/(2\pi)^{3} v_{x} f_{\nu}$ 
satisfy [\onlinecite{valet93}] 
\begin{equation}
  \frac{\partial n_{\nu}}{\partial t}
  +
  \frac{\partial j_{\nu}}{\partial x}
  =
  -\frac{n_{\nu}}{2 \tau_{\rm sf}^{\nu}}
  +
  \frac{n_{-\nu}}{2 \tau_{\rm sf}^{-\nu}}, 
  \label{eq:continuous_equation_ns}
\end{equation}
where the spin-flip scattering time from spin state $\nu$ to $-\nu$ (up to down or down to up) is 
denoted by $\tau_{\rm sf}^{\nu}$. 
The charge density $n_{e}=-e(n_{\uparrow}+n_{\downarrow})$ and 
electric current density $J_{e}=-e(j_{\uparrow}+j_{\downarrow})$ 
satisfy the conservation law, $\partial n_{e}/\partial t + \partial J_{e}/\partial x=0$. 
The electron density $n_{\nu}$ is related to $\delta\mu_{\nu}$ via $n_{\nu} \simeq \mathcal{N}_{\nu} \delta\mu_{\nu}$, 
where $\mathcal{N}_{\nu}$ is the density of states of the spin-$\nu$ electron at the Fermi level. 
In the diffusive metal, 
$j_{\nu}$ can be expressed as 
\begin{equation}
  j_{\nu}
  =
  -\frac{\sigma_{\nu}}{e^{2}}
  \frac{\partial \bar{\mu}_{\nu}}{\partial x}, 
  \label{eq:longitudinal_current}
\end{equation}
where the conductivity of the spin-$\nu$ electron $\sigma_{\nu}$ relates to 
the diffusion constant $D_{\nu}$ and the density of state $\mathcal{N}_{\nu}$ via 
the Einstein law $\sigma_{\nu}=e^{2}\mathcal{N}_{\nu}D_{\nu}$. 
Detailed balance [\onlinecite{hershfield97}], 
$\mathcal{N}_{\uparrow}/\tau_{\rm sf}^{\uparrow}=\mathcal{N}_{\downarrow}/\tau_{\rm sf}^{\downarrow}$, is 
satisfied in the steady state. 
The spin polarizations of the conductivity and the diffusion constant are denoted by 
$\beta=(\sigma_{\uparrow}-\sigma_{\downarrow})/(\sigma_{\uparrow}+\sigma_{\downarrow})$ 
and $\beta^{\prime}=(D_{\uparrow}-D_{\downarrow})/(D_{\uparrow}+D_{\downarrow})$. 
From Eq. (\ref{eq:continuous_equation_ns}), 
the longitudinal spin accumulation in the steady state satisfies 
the diffusion equation [\onlinecite{valet93}] 
\begin{equation}
  \frac{\partial^{2}}{\partial x^{2}}
  \bm{\mu}^{\rm L}
  = 
  \frac{1}{\lambda_{\rm sd(L)}^{2}}
  \bm{\mu}^{\rm L}, 
  \label{eq:diffusion_equation_longitudinal}
\end{equation}
where $\lambda_{\rm sd(L)}$ is the longitudinal spin diffusion length 
defined as $1/\lambda_{\rm sd(L)}^{2}=[1/(D_{\uparrow}\tau_{\rm sf}^{\uparrow})+1/(D_{\downarrow}\tau_{\rm sf}^{\downarrow})]/2$. 
The longitudinal spin current density can be expressed as 
\begin{equation}
  \mathbf{J}_{s}^{\rm L}
  =
  -\frac{\hbar}{2e^{2}}
  \frac{\partial}{\partial x}
  \left(
    \sigma_{\uparrow}
    \bar{\mu}_{\uparrow}
    -
    \sigma_{\downarrow}
    \bar{\mu}_{\downarrow}
  \right)
  \mathbf{m}. 
  \label{eq:longitudinal_spin_current}
\end{equation}


The issue of whether transport of the transverse spin 
in the ferromagnet is 
ballistic or diffusive has been discussed in 
[\onlinecite{slonczewski96,stiles02,brataas06}] and [\onlinecite{zhang02,shpiro03,zhang04,piechon07}]. 
These two theories are supported by different experiments [\onlinecite{taniguchi08a,chen06,taniguchi08c,ghosh12}], 
and the validity of each theory is still controversial. 
The present work considers the case of diffusive transport for generality. 
Ballistic transport corresponds to the limit of $\lambda_{J}, t_{\rm r(i)}^{\uparrow\downarrow} \to 0$, 
where $\lambda_{J}$ is the spin coherence length introduced below. 
In the steady state, 
the transverse spin accumulation $\bm{\mu}^{\rm T}=\bm{\mu}-\bm{\mu}^{\rm L}$ obeys [\onlinecite{zhang02,taniguchi08a}] 
\begin{equation}
  \frac{\partial^{2}}{\partial x^{2}}
  \bm{\mu}^{\rm T}
  =
  \frac{1}{\lambda_{J}^{2}}
  \bm{\mu}^{\rm T}
  \times
  \mathbf{m}
  +
  \frac{1}{\lambda_{\rm sd(T)}^{2}}
  \bm{\mu}^{\rm T}, 
  \label{eq:diffusion_equation_transverse}
\end{equation}
where the first term on the right-hand-side describes 
the precession of the spin accumulation around the magnetization 
due to the exchange coupling. 
The exchange coupling constant $J_{\rm sd}$ is in relation to the spin coherence length $\lambda_{J}$ via 
$\lambda_{J}=\sqrt{\hbar(D_{\uparrow}+D_{\downarrow})/(2J_{\rm sd})}$ [\onlinecite{simanek01,zhang02,shpiro03,zhang04,piechon07,taniguchi09}]. 
The spin diffusion length of the transverse spin is $\lambda_{\rm sd(T)}$ [\onlinecite{zhang02}]. 
The transverse spin current density is related to 
the transverse spin accumulation via [\onlinecite{taniguchi08a,zhang02}] 
\begin{equation}
  \mathbf{J}_{s}^{\rm T}
  =
  -\frac{\hbar \sigma_{\uparrow\downarrow}}{2e^{2}}
  \frac{\partial}{\partial x}
  \bm{\mu}^{\rm T}, 
  \label{eq:transverse_spin_current}
\end{equation}
where 
$\sigma_{\uparrow\downarrow}=e^{2}[(\mathcal{N}_{\uparrow}+\mathcal{N}_{\downarrow})/2][(D_{\uparrow}+D_{\downarrow})/2]$. 
The solutions of the transverse spin accumulation and current are linear combinations of 
$e^{\pm x/\ell}$ and $e^{\pm x/\ell^{*}}$ 
with $1/\ell=\sqrt{(1/\lambda_{\rm sd(T)}^{2}) - (i/\lambda_{J}^{2})}$. 


In the nonmagnetic layer, 
the distinction between the longitudinal and transverse spin is unnecessary. 
In fact, in the limit of zero-spin polarization ($\beta=\beta^{\prime}=0$) and 
in the absence of the exchange coupling between the magnetization and electrons' spin ($J_{\rm sd}=0$), 
as for the nonmagnet, 
Eqs. (\ref{eq:diffusion_equation_longitudinal}) and (\ref{eq:diffusion_equation_transverse}), 
or Eqs. (\ref{eq:longitudinal_spin_current}) and (\ref{eq:transverse_spin_current}), 
become identical. 


The relation between the spin accumulation and dissipation is as follows. 
The heat density of the longitudinal spin-$\nu$ electrons $dq_{\nu}$ relates to 
the energy density $u_{\nu}=\int d^{3}\mathbf{k}/(2\pi)^{3} \varepsilon f_{\nu}$, 
chemical potential $\mu_{\nu}=\mu_{0}+\delta\mu_{\nu}$, 
and the electron density $n_{\nu}$ via [\onlinecite{ashcroft76,kondepudi98}] 
\begin{equation}
  dq_{\nu}
  =
  du_{\nu}
  -\mu_{\nu} 
  dn_{\nu}. 
  \label{eq:thermodynamic_relation}
\end{equation}
The energy density $u^{\rm L}=u_{\uparrow}+u_{\downarrow}$ 
for the longitudinal spin satisfies [\onlinecite{rammer08}] 
\begin{equation}
  \frac{\partial u^{\rm L}}{\partial t}
  +
  \frac{\partial j_{u}^{\rm L}}{\partial x}
  =
  J_{e}E,
\end{equation}
where $j_{u}^{\rm L}=j_{u,\uparrow}+j_{u,\downarrow}$, 
and $j_{u,\nu}=\int d^{3}\mathbf{k}/(2\pi)^{3} \varepsilon v_{x} f_{\nu}$ 
is the energy current density [\onlinecite{rammer08}]. 
Here, the term $J_{e}E$ is the Joule heating due to the electric current. 
On the other hand, the energy current of the transverse spin $j_{u}^{\rm T}$ 
satisfies $\partial j_{u}^{\rm T}/\partial x=0$ in the steady state, 
where the right-hand-side is zero because there is no source of the transverse spin 
inside the F and N layers. 
We introduce the heat current density by [\onlinecite{comment1}] 
\begin{equation}
  j_{q}
  =
  j_{u}^{\rm L}
  -
  \sum_{\nu=\uparrow,\downarrow}
  \mu_{\nu}
  j_{\nu}
  +
  j_{u}^{\rm T}
  -
  \bm{\mu}^{\rm T}
  \cdot
  \frac{\mathbf{J}_{s}^{\rm T}}{\hbar}. 
  \label{eq:heat_current}
\end{equation}
In steady state, the heat current is related to the dissipation via [\onlinecite{ziman07}] 
$\partial Q_{V}/\partial t=T[\partial (j_{q}/T)/\partial x]$, 
where the temperature $T$ is assumed to be spatially uniform in the following calculations. 
The subscript "$V$" is used to emphasize that 
this is the dissipation per unit volume per unit time. 
Then, $\partial Q_{V}/\partial t$ is 
\begin{equation}
  \frac{\partial Q_{V}}{\partial t}
  =
  \frac{J_{e}}{e}
  \frac{\partial\bar{\mu}}{\partial x}
  -
  \frac{\partial}{\partial x}
  \frac{\mathbf{J}_{s}}{\hbar}
  \cdot
  \bm{\mu},
  \label{eq:dissipation_volume} 
\end{equation}
where $\bar{\mu}=(\bar{\mu}_{\uparrow}+\bar{\mu}_{\downarrow})/2$ is the electrochemical potential. 
The interface resistance also gives the dissipation, 
where the dissipation per unit area per unit time is 
\begin{equation}
  \frac{\partial Q_{A}}{\partial t}
  =
  \frac{J_{e}}{e}
  \delta
  \bar{\mu}
  -
  \frac{\mathbf{J}_{s}}{\hbar}
  \cdot
  \delta
  \bm{\mu},
  \label{eq:dissipation_area}
\end{equation}
where $\delta\bar{\mu}$ and $\delta\bm{\mu}$ are 
the differences of $\bar{\mu}$ and $\bm{\mu}$ at the F/N interface. 
The subscript "$A$" is used to emphasize that 
this is the dissipation per unit area per unit time. 
Equations (\ref{eq:dissipation_volume}) and (\ref{eq:dissipation_area}) are 
generalized Joule heating formulas 
in the presence of spin current, 
and the main results in this section. 
The total spin current $\mathbf{J}_{s}$ and spin accumulation $\bm{\mu}$ include 
both the longitudinal and transverse components, 
whereas only the longitudinal components appeared in the previous work [\onlinecite{tulapurkar11}]. 
The amount of the dissipation can be evaluated by 
substituting the solution of the diffusion equation of the spin accumulation 
into Eqs. (\ref{eq:dissipation_volume}) and (\ref{eq:dissipation_area}) 
with accurate boundary conditions provided by 
Eqs. (\ref{eq:spin_pumping}) and (\ref{eq:backflow}). 
We call Eqs. (\ref{eq:dissipation_volume}) and (\ref{eq:dissipation_area}) 
the bulk and interface dissipations, respectively. 



\section{Dissipation due to spin pumping}
\label{sec:Dissipation due to spin pumping}

In spin pumping, 
transverse spin angular momentum is steadily transferred 
from the magnetic system (F${}_{1}$ layer) to the conduction electrons near the F${}_{1}$/N interface. 
The net spin angular momentum, 
$d \mathbf{s}=[\mathbf{I}_{s}^{\rm pump}+\mathbf{m}_{1}\times(\mathbf{I}_{s}^{\rm F_{1} \to N} \times \mathbf{m}_{1})]dt$, 
transferred from the ferromagnet should overcome 
the potential difference $\bm{\mu}_{\rm N}-\bm{\mu}_{\rm F_{1}}$ 
to be pumped steadily from the F${}_{1}$/N interface to the N layer during the time $dt$. 
This means that not only the spin angular momentum 
but also the energy is transferred 
from the F${}_{1}$ layer to the conduction electrons. 
The transferred energy per unit area per unit time is given by 
$(\bm{\mu}_{\rm N}-\bm{\mu}_{\rm F_{1}})\cdot(d\mathbf{s}/dt)/(\hbar S)$. 
In terms of the spin current and spin accumulation, 
this transferred energy is expressed as 
\begin{equation}
\begin{split}
  \frac{\partial Q_{A}^{\rm SP}}{\partial t}
  =&
  \frac{1}{\hbar S}
  \left[
    \mathbf{I}_{s}^{\rm pump}
    +
    \mathbf{m}_{1}
    \times
    \left(
      \mathbf{I}_{s}^{\rm F_{1} \to N}
      \times
      \mathbf{m}_{1}
    \right)
  \right]
\\
  &\cdot
  \left[
    \bm{\mu}_{\rm N}(x=0)
    -
    \bm{\mu}_{\rm F_{1}}(x=0)
  \right].
  \label{eq:dissipation_FNF}
\end{split}
\end{equation}
Comparing Eq. (\ref{eq:dissipation_FNF}) with Eq. (\ref{eq:dissipation_area}), 
we find the relation 
\begin{equation}
  \left(
    \frac{\partial Q_{A}}{\partial t}
  \right)_{\rm F_{1}/N}^{\rm T} 
  =
  -\frac{\partial Q_{A}^{\rm SP}}{\partial t}, 
\end{equation}
where $(\partial Q_{A}/\partial t)_{\rm F_{1}/N}^{\rm T}$ is defined by 
\begin{equation}
  \left(
    \frac{\partial Q_{A}}{\partial t}
  \right)_{\rm F_{1}/N}^{\rm T} 
  =
  \left(
    \frac{\partial Q_{A}}{\partial t}
  \right)_{\rm F_{1}/N}
  -
  \left(
    \frac{\partial Q_{A}}{\partial t}
  \right)_{\rm F_{1}/N}^{\rm L}. 
  \label{eq:transverse_interface_dissipation_F1N}
\end{equation}
Here, $(\partial Q_{A}/\partial t)_{\rm F_{1}/N}$ is 
the F${}_{1}$/N interface dissipation defined by Eq. (\ref{eq:dissipation_area}), 
whereas 
\begin{equation}
\begin{split}
  \left(
    \frac{\partial Q_{A}}{\partial t}
  \right)_{\rm F_{1}/N}^{\rm L}
  =&
  -\frac{1}{\hbar S}
  \left(
    \mathbf{m}_{1}
    \cdot
    \mathbf{I}_{s}^{\rm F_{1} \to N}
  \right)
  \mathbf{m}_{1}
\\
  &\cdot
  \left[
    \bm{\mu}_{\rm N}(x=0)
    -
    \bm{\mu}_{\rm F_{1}}(x=0)
  \right]. 
  \label{eq:longitudinal_interface_dissipation_F1N}
\end{split}
\end{equation}
Because Eq. (\ref{eq:longitudinal_interface_dissipation_F1N}) 
is defined by the longitudinal components of the spin current and spin accumulation in Eq. (\ref{eq:dissipation_area}), 
we call this quantity the longitudinal part of the F${}_{1}$/N interface dissipation. 
On the other hand, Eq. (\ref{eq:transverse_interface_dissipation_F1N}) is defined 
by the transverse components of the spin current and spin accumulation at the F${}_{1}$/N interface. 
Moreover, using Eqs. (\ref{eq:dissipation_volume}), (\ref{eq:dissipation_area}) and (\ref{eq:transverse_interface_dissipation_F1N}), 
Eq. (\ref{eq:dissipation_FNF}) can be rewritten as 
\begin{equation}
\begin{split}
  \frac{\partial Q_{A}^{\rm SP}}{\partial t}
  =&
  \left(
    \frac{\partial Q_{A}}{\partial t}
  \right)_{\rm F_{2}/N}
  +
  \int_{0}^{d_{2}} dx 
  \left(
    \frac{\partial Q_{V}}{\partial t}
  \right)_{\rm F_{2}} 
\\
  &+
  \left(
    \frac{\partial Q_{A}}{\partial t}
  \right)_{\rm F_{1}/N}^{\rm L}
  +
  \int_{-d_{1}}^{0} dx 
  \left(
    \frac{\partial Q_{V}}{\partial t}
  \right)_{\rm F_{1}},
   \label{eq:dissipation_relation}
\end{split}
\end{equation}
where the F${}_{2}$/N interface dissipation, $(\partial Q_{A}/\partial t)_{\rm F_{2}/N}$ in Eq. (\ref{eq:dissipation_relation}), 
and the F${}_{1}$ and F${}_{2}$ bulk dissipations, 
$(\partial Q_{V}/\partial t)_{\rm F_{1}}$ and $(\partial Q_{V}/\partial t)_{\rm F_{2}}$, 
are defined from Eqs. (\ref{eq:dissipation_volume}) and (\ref{eq:dissipation_area}). 
As discussed below, 
Eq. (\ref{eq:dissipation_relation}) describes the energy dissipation process 
carried by the spin current. 
Therefore, we define Eq. (\ref{eq:dissipation_relation}), 
or equivalently, Eq. (\ref{eq:dissipation_FNF}), the dissipation due to spin pumping. 



\begin{figure}
\centerline{\includegraphics[width=0.8\columnwidth]{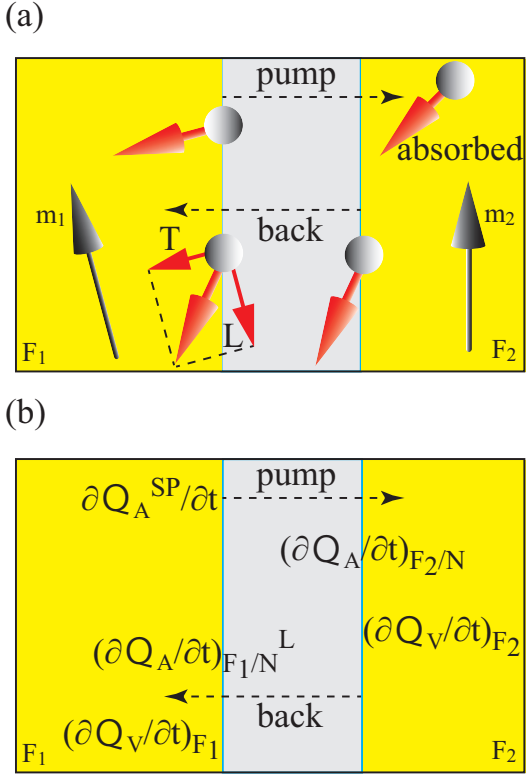}}\vspace{-3.0ex}
\caption{
         Schematic views of the flows of (a) angular momentum and (b) energy 
         from the microwave to the ferromagnetic multilayer, 
         in which "L" and "T" define the longitudinal and transverse components 
         with respect to $\mathbf{m}_{1}$. 
         \vspace{-3ex}}
\label{fig:fig2}
\end{figure}




With the help of Figs. \ref{fig:fig2} (a) and \ref{fig:fig2} (b) 
we now discuss the physical interpretation of Eq. (\ref{eq:dissipation_relation}), 
which schematically show the flows of spin angular momentum and of energy. 
In spin pumping one usually focuses  
attention only on the flow of spin angular momentum, i.e., spin current, but 
because we are also interested in energy dissipation we also show energy flow. 
When the pumped angular momentum reaches the F${}_{2}$/N interface, 
part of it is absorbed in the F${}_{2}$ layer, 
and is depolarized by scattering at the F${}_{2}$/N interface 
and by spin flip and spin diffusion within the F${}_{2}$ layer. 
The remaining part returns to the F${}_{1}$/N interface, which we call back flow. 
The back flow to the F${}_{1}$ layer is relaxed 
by scattering at the F${}_{1}$/N interface and by spin flip and spin diffusion within the F${}_{1}$ layer, 
where the transverse component of the back flow at the F${}_{1}$/N interface 
renormalizes the pumped spin current. 
In terms of the energy flow shown in Fig. \ref{fig:fig2} (b), 
spin absorption at the F${}_{2}$/N interface leads to 
the interface dissipation $(\partial Q_{A}/\partial t)_{\rm F_{2}/N}$ 
and bulk dissipation $(\partial Q_{V}/\partial t)_{\rm F_{2}}$ 
due to spin depolarization.
The back flow at the F${}_{1}$ layer also gives 
the interface dissipation $(\partial Q_{A}/\partial t)_{\rm F_{1}/N}^{\rm L}$ 
and bulk dissipation $(\partial Q_{V}/\partial t)_{\rm F_{1}}$. 
The total dissipation is the sum of these dissipations, as indicated by Eq. (\ref{eq:dissipation_relation}). 
In other words, the transferred energy from the F${}_{1}$ layer to the conduction electrons 
at the F${}_{1}$/N interface is not localized, 
and is dissipated throughout the system. 
Then, Eq. (\ref{eq:dissipation_relation}), or equivalently, Eq. (\ref{eq:dissipation_FNF}), 
can be regarded as the dissipation due to spin pumping. 
Also, Eq. (\ref{eq:transverse_interface_dissipation_F1N}) is regarded as 
the energy transfer from the F${}_{1}$ layer to 
the conduction electrons near the F${}_{1}$/N interface. 
Appendix A shows that all terms on the right-hand side of Eq. (\ref{eq:dissipation_relation}) are positive, 
thus guaranteeing the second law of thermodynamics. 


To conclude this section, 
it is of interest to compare Eq. (\ref{eq:dissipation_FNF}) with 
the dissipation due to electric current. 
Let us assume that an electric current is flowing through a multilayer, 
driven by a voltage difference across two electrodes. 
The total dissipation per unit area per unit time is obtained 
from Eqs. (\ref{eq:dissipation_volume}) and (\ref{eq:dissipation_area}) as [\onlinecite{tulapurkar11}] 
\begin{equation}
  \frac{\partial Q_{A}^{\rm EC}}{\partial t}
  =
  \frac{J_{e}}{e}
  \left[
    \bar{\mu}(\infty)
    -
    \bar{\mu}(-\infty)
  \right],
  \label{eq:dissipation_electric_current}
\end{equation}
where $[\bar{\mu}(\infty)-\bar{\mu}(-\infty)]/e$ is the voltage difference between the electrodes. 
Comparing Eq. (\ref{eq:dissipation_FNF}) with (\ref{eq:dissipation_electric_current}), 
we notice that 
the net transverse spin current and the difference in the spin accumulation at the F${}_{1}$/N interface 
correspond to the electric current and applied voltage, respectively, 
and that in spin pumping the F${}_{1}$/N interface plays the role of the electrode, 
This is because the angular momentum and the energy 
transferred from the magnetization of the F${}_{1}$ layer to 
the conduction electron are pumped from this interface to the multilayer. 



\section{Evaluation of dissipation}
\label{sec:Evaluation of dissipation}

In this section, we quantitatively evaluate 
the dissipation due to spin pumping, Eq. (\ref{eq:dissipation_FNF}). 
Substituting the solutions of Eqs. (\ref{eq:diffusion_equation_longitudinal}) and (\ref{eq:diffusion_equation_transverse}) 
into Eq. (\ref{eq:backflow}), 
the total spin currents at the F${}_{1}$/N and F${}_{2}$/N interfaces are, respectively, expressed as 
\begin{equation}
\begin{split}
  &
  \mathbf{I}_{s}^{\rm pump}
  +
  \mathbf{I}_{s}^{\rm F_{1} \to N}
  =
  \frac{\hbar}{4\pi}
  \left(
    \tilde{g}_{\rm r(F_{1})}^{\uparrow\downarrow}
    \mathbf{m}_{1}
    \times
    \frac{d\mathbf{m}_{1}}{dt}
    +
    \tilde{g}_{\rm i(F_{1})}^{\uparrow\downarrow}
    \frac{d\mathbf{m}_{1}}{dt}
  \right)
\\
  &-
  \frac{1}{4\pi}
  \left[
    \tilde{g}_{\rm F_{1}}^{*}
    \left(
      \mathbf{m}_{1}
      \cdot
      \bm{\mu}_{\rm N}
    \right)
    \mathbf{m}_{1}
    +
    \tilde{g}_{\rm r(F_{1})}^{\uparrow\downarrow}
    \mathbf{m}_{1}
    \times
    \left(
      \bm{\mu}_{\rm N}
      \times
      \mathbf{m}_{1}
    \right)
  \right.
\\
  &
  \left.
  \ \ \ \ \ \ +
    \tilde{g}_{\rm i(F_{1})}^{\uparrow\downarrow}
    \bm{\mu}_{\rm N}
    \times
    \mathbf{m}_{1}
  \right],
  \label{eq:spin_current_F1N_renormalized}
\end{split}
\end{equation}
\begin{equation}
\begin{split}
  \mathbf{I}_{s}^{\rm F_{2} \to N}
  =
  -\frac{1}{4\pi}
  &
  \left[
    \tilde{g}_{\rm F_{2}}^{*}
    \left(
      \mathbf{m}_{2}
      \cdot
      \bm{\mu}_{\rm N}
    \right)
    \mathbf{m}_{2}
    +
    \tilde{g}_{\rm r(F_{2})}^{\uparrow\downarrow}
    \mathbf{m}_{2}
    \times
    \left(
      \bm{\mu}_{\rm N}
      \times
      \mathbf{m}_{2}
    \right)
  \right.
\\
  &
  \left.
  \ \ \ \ \ \ +
    \tilde{g}_{\rm i(F_{2})}^{\uparrow\downarrow}
    \bm{\mu}_{\rm N}
    \times
    \mathbf{m}_{2}
  \right].
  \label{eq:spin_current_F2N_renormalized}
\end{split}
\end{equation}
The renormalized conductances, $\tilde{g}^{*}$ and $\tilde{g}_{\rm r,i}^{\uparrow\downarrow}$, 
are defined by the following ways: 
\begin{equation}
  \frac{1}{\tilde{g}^{*}}
  =
  \frac{2}{(1-\gamma^{2})g} 
  +
  \frac{1}{g_{\rm sd} \tanh(d/\lambda_{\rm sd(L)})},
  \label{eq:g_star}
\end{equation}
\begin{equation}
  \begin{pmatrix}
    \tilde{g}_{\rm r}^{\uparrow\downarrow} \\
    \tilde{g}_{\rm i}^{\uparrow\downarrow} 
  \end{pmatrix}
  =
  \frac{1}{K_{1}^{2}+K_{2}^{2}}
  \begin{pmatrix}
    K_{1} & K_{2} \\
    -K_{2} & K_{1}
  \end{pmatrix}
  \begin{pmatrix}
    g_{\rm r}^{\uparrow\downarrow} \\
    g_{\rm i}^{\uparrow\downarrow}
  \end{pmatrix}, 
\end{equation}
where $g_{\rm sd}=h(1-\beta^{2})S/(2e^{2}\rho \lambda_{\rm sd(L)})$, 
and $\rho=1/(\sigma^{\uparrow}+\sigma^{\downarrow})$ is the resistivity. 
The terms $K_{1}$ and $K_{2}$ are defined as 
\begin{equation}
  K_{1}
  =
  1
  +
  t_{\rm r}^{\uparrow\downarrow}
  {\rm Re}
  \left[
    \frac{1}{g_{\rm t}\tanh(d/\ell)}
  \right]
  +
  t_{\rm i}^{\uparrow\downarrow}
  {\rm Im}
  \left[
    \frac{1}{g_{\rm t}\tanh(d/\ell)}
  \right],
\end{equation}
\begin{equation}
  K_{2}
  =
  t_{\rm i}^{\uparrow\downarrow}
  {\rm Re}
  \left[
    \frac{1}{g_{\rm t}\tanh(d/\ell)}
  \right]
  -
  t_{\rm r}^{\uparrow\downarrow}
  {\rm Im}
  \left[
    \frac{1}{g_{\rm t}\tanh(d/\ell)}
  \right],
\end{equation}
where $g_{\rm t}=hS \sigma_{\uparrow\downarrow}/(e^{2}\ell)$. 
In the ballistic transport limit for the transverse spin, 
$\tilde{g}^{\uparrow\downarrow}$ equals to $g^{\uparrow\downarrow}$. 
Then, we expand $\bm{\mu}_{\rm N}$ as 
$\bm{\mu}_{\rm N}=\hbar( \omega a \sin\theta \mathbf{m}_{1} + b \dot{\mathbf{m}}_{1} + c \mathbf{m}_{1} \times \dot{\mathbf{m}}_{1})$, 
where $y=\delta_{y}/\Delta$ ($y=a,b,c$) are dimensionless coefficients 
determined by Eq. (\ref{eq:conservation_spin_current}) with 
Eqs. (\ref{eq:spin_current_F1N_renormalized}) and (\ref{eq:spin_current_F2N_renormalized}). 
In the limit of $g_{\rm r}^{\uparrow\downarrow} \gg g_{\rm i}^{\uparrow\downarrow}$ [\onlinecite{brataas06}], 
$\delta_{b} =0$, 
and $\Delta$, $\delta_{a}$, and $\delta_{c}$ are given by 
\begin{equation}
\begin{split}
  \Delta 
  =&
  \left(
    \tilde{g}_{\rm r(F_{1})}^{\uparrow\downarrow}
    +
    \tilde{g}_{\rm r(F_{2})}^{\uparrow\downarrow}
  \right)
  \left[
    \left(
      \tilde{g}_{\rm F_{1}}^{*}
      +
      \tilde{g}_{\rm F_{2}}^{*}
      \cos^{2}\theta
      +
      \tilde{g}_{\rm r(F_{2})}^{\uparrow\downarrow}
      \sin^{2}\theta
    \right)
  \right.
\\
  &
  \left. 
  \times
    \left(
      \tilde{g}_{\rm r(F_{1})}^{\uparrow\downarrow}
      +
      \tilde{g}_{\rm r(F_{2})}^{\uparrow\downarrow}
      \cos^{2}\theta
      +
      \tilde{g}_{\rm F_{2}}^{*}
      \sin^{2}\theta
    \right)
  \right. 
\\
  &
  \left. 
    -
    \left(
      \tilde{g}_{\rm r(F_{2})}^{\uparrow\downarrow}
      -
      \tilde{g}_{\rm F_{2}}^{*}
    \right)^{2}
    \sin^{2}\theta
    \cos^{2}\theta
  \right],
  \label{eq:Delta}
\end{split}
\end{equation}
\begin{equation}
  \delta_{a}
  =
  \tilde{g}_{\rm r(F_{1})}^{\uparrow\downarrow}
  \left(
    \tilde{g}_{\rm r(F_{1})}^{\uparrow\downarrow}
    +
    \tilde{g}_{\rm r(F_{2})}^{\uparrow\downarrow}
  \right)
  \left(
    \tilde{g}_{\rm r(F_{2})}^{\uparrow\downarrow}
    -
    \tilde{g}_{\rm F_{2}}^{*}
  \right)
  \sin\theta
  \cos\theta,
  \label{eq:delta_a}
\end{equation}
\begin{equation}
\begin{split}
  \delta_{c}
  =&
  \tilde{g}_{\rm r(F_{1})}^{\uparrow\downarrow}
  \left(
    \tilde{g}_{\rm r(F_{1})}^{\uparrow\downarrow}
    +
    \tilde{g}_{\rm r(F_{2})}^{\uparrow\downarrow}
  \right)
\\
  & \times 
  \left(
    \tilde{g}_{\rm F_{1}}^{*}
    +
    \tilde{g}_{\rm F_{2}}^{*}
    \cos^{2}\theta
    +
    \tilde{g}_{\rm r(F_{2})}^{\uparrow\downarrow}
    \sin^{2}\theta
  \right).
  \label{eq:delta_c}
\end{split}
\end{equation}



Equation (\ref{eq:dissipation_FNF}) 
in the limit of $g_{\rm r}^{\uparrow\downarrow} \gg g_{\rm i}^{\uparrow\downarrow}$ is 
then given by 
\begin{equation}
\begin{split}
  \frac{\partial Q_{A}^{\rm SP}}{\partial t}
  =
&
  \frac{\hbar \omega^{2} \sin^{2}\theta \tilde{g}_{\rm r(F_{1})}^{\uparrow\downarrow}(1-c)}{4\pi S}
\\
& \times
  \left\{
    c
    +
    \tilde{g}_{\rm r(F_{1})}^{\uparrow\downarrow}
    \left(
      1
      -
      c
    \right)
    {\rm Re}
    \left[
      \frac{1}{g_{\rm t}\tanh(d_{1}/\ell)}
    \right]
  \right\}.
  \label{eq:dissipation_quantitative}
\end{split}
\end{equation}
In the ballistic transport limit of the transverse spin, 
Eq. (\ref{eq:dissipation_quantitative}) is simplified to 
$\hbar \omega^{2} g_{\rm r(F_{1})}^{\uparrow\downarrow}(1-c)c/(4\pi S)$. 
We emphasize that Eq. (\ref{eq:dissipation_quantitative}) is proportional to 
the enhancement of the Gilbert damping by spin pumping [\onlinecite{tserkovnyak03,taniguchi08a}]: 
\begin{equation}
  \alpha^{\prime}
  =
  \frac{\gamma_{0}\hbar \tilde{g}_{\rm r(F_{1})}^{\uparrow\downarrow}(1-c)}{4\pi MSd_{1}},
  \label{eq:alpha}
\end{equation}
where $\gamma_{0}$ is the gyromagnetic ratio. 
Here, $\alpha^{\prime}$ is derived in the following way. 
According to the conservation law of the total angular momentum, 
the pumped spin from the F${}_{1}$/N interface per unit time, $d \mathbf{s}/dt$, 
should equal to the time change of the magnetization in the F${}_{1}$ layer, 
i.e., a torque $d \mathbf{m}_{1}/dt=[(g \mu_{\rm B})/(\hbar MSd)]d \mathbf{s}/dt$ acts on $\mathbf{m}_{1}$, 
where $M/(g \mu_{\rm B})$ is the number of the magnetic moments in the F${}_{1}$ layer, 
and the Land\'e $g$-factor satisfies $g \mu_{\rm B}=\gamma_{0}\hbar$. 
This torque, $[(g \mu_{\rm B})/(\hbar MSd)]d \mathbf{s}/dt$, with 
$d \mathbf{s}/dt=\mathbf{I}_{s}^{\rm pump}+\mathbf{m}_{1} \times (\mathbf{I}_{s}^{\rm F_{1} \to N} \times \mathbf{m}_{1})$, 
can be expressed as $\alpha^{\prime} \mathbf{m}_{1} \times (d \mathbf{m}_{1}/dt)$. 
Then, $\alpha^{\prime}$ is identified as the enhancement of the Gilbert damping constant 
due to the spin pumping. 
The present result indicating that the dissipation is proportional to $\alpha^{\prime}$ 
represents that the pumped spin current at the F${}_{1}$/N interface carries 
not only the angular momentum but also the energy 
from the F${}_{1}$ to N layer. 



\begin{figure}
\centerline{\includegraphics[width=0.8\columnwidth]{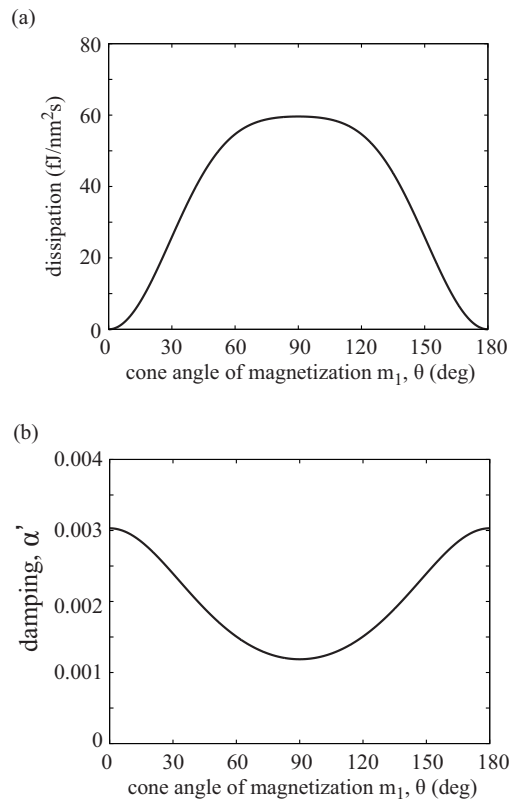}}\vspace{-3.0ex}
\caption{
         Dependencies of 
         (a) the dissipation due to pure spin-current, Eq. (\ref{eq:dissipation_FNF}), 
         and (b) the damping, $\alpha^{\prime}$, Eq. (\ref{eq:alpha}), on the cone angle $\theta$. 
         \vspace{-3ex}}
\label{fig:fig3}
\end{figure}



We quantitatively evaluate Eq. (\ref{eq:dissipation_quantitative}) 
by using parameters taken from experiments for the NiFe/Cu multilayer 
with the assumption $\beta=\beta^{\prime}$ [\onlinecite{zhang02,taniguchi08a,bass07,fert99}]; 
$(h/e^{2})S/[(1-\gamma^{2})g]=0.54$ k$\Omega$nm${}^{2}$, 
$\gamma=0.7$, 
$g_{\rm r}^{\uparrow\downarrow}/S=15$ nm${}^{-2}$, 
$g_{\rm i}^{\uparrow\downarrow}/S=1$ nm${}^{-2}$, 
$t_{\rm r}^{\uparrow\downarrow}/S=t_{\rm i}^{\uparrow\downarrow}/S=4$ nm${}^{-2}$, 
$\rho=241$ $\Omega$nm, 
$\beta=0.73$, 
$\lambda_{\rm sd(L)}=5.5$ nm, 
$\lambda_{\rm sd(T)}=\lambda_{\rm sd(L)}/\sqrt{1-\beta^{2}}$,
$\lambda_{J}=2.8$ nm, 
$d=5$ nm, 
$\gamma_{0}=1.8467 \times 10^{11}$ rad/(T s), 
$M=605 \times 10^{3}$ A/m, 
and $\omega=2\pi \times 9.4 \times 10^{9}$ rad/s, respectively, 
where the parameters of the F${}_{1}$ and F${}_{2}$ layers are assumed to be identical, for simplicity. 
In Fig. \ref{fig:fig3} (a), 
we show the dissipation due to spin pumping, Eq. (\ref{eq:dissipation_quantitative}), 
for an arbitrary cone angle $\theta$. 
The damping $\alpha^{\prime}$, Eq. (\ref{eq:alpha}), is also shown in Fig. \ref{fig:fig3} (b). 
The cone angle $\theta$ in typical FMR experiments [\onlinecite{mizukami02a,mizukami02b}] is small. 
However, the spin pumping affects not only the FMR experiment 
but also spin torque switching [\onlinecite{chen06}], 
in which $\theta$ varies from $0^{\circ}$ to $180^{\circ}$. 
Therefore, we show the dissipation and damping for the whole range of $\theta$ in Fig. \ref{fig:fig3}. 


The dissipation is zero for $\theta=0^{\circ}$ and $180^{\circ}$ 
because $d \mathbf{m}_{1}/dt=\bm{0}$ at these angles. 
The maximum dissipation is about 60 fJ/(nm${}^{2}$s). 
To understand how large this dissipation is, 
we compare this value 
with the dissipation due to spin torque switching current in the same system;  
we discuss this in the next section. 


To conclude this section, 
we briefly mention that the dissipation due to spin pumping 
can be evaluated not only from Eq. (\ref{eq:dissipation_FNF}) 
but also from Eq. (\ref{eq:dissipation_relation}). 
Appendix B gives explicit forms for each term 
on the right-hand side of Eq. (\ref{eq:dissipation_relation}), 
from which the dissipation can be calculated. 



\section{Comparison with spin torque switching}
\label{sec:Comparison with spin torque switching}

Spin pumping occurs not only in FMR experiments 
but also in spin torque switching experiments. 
An important issue in the spin torque switching problem is 
the reduction of power consumption due to heating [\onlinecite{locatelli13}]. 
Whereas heating has usually meant the dissipation due to electric current, 
the results of the previous section indicate that spin pumping also contributes to the dissipation. 
Thus it is of interest to quantitatively evaluate the dissipation due to the electric current, 
and compare it with that due to spin pumping studied in the previous section, 
which will clarify the ratio of the contribution of spin pumping 
to heating in the spin torque switching experiment. 


We assume that an electric current $I$ is injected from the F${}_{2}$ layer to the F${}_{1}$ layer. 
Then, a term 
\begin{equation}
  \mathbf{I}_{s(e)}^{{\rm F}_{k} \to {\rm N}}
  =
  \frac{\hbar \gamma}{2e}
  I^{{\rm F}_{k} \to {\rm N}}
  \mathbf{m}_{k},
  \label{eq:backflow_electric}
\end{equation}
should be added to Eq. (\ref{eq:backflow}), 
which represents a spin current due to the electric current [\onlinecite{brataas06}]. 
The current $I^{{\rm F}_{k} \to N}$ is the electric current 
which flows from the F${}_{k}$ layer to the N layer, 
meaning that $I^{{\rm F}_{1} \to N}=-I^{{\rm F}_{2} \to N}=-I$. 
As in the system studied in the previous section, 
we assume that the spin current is zero at both ends of the ferromagnet. 
Taking into account Eq. (\ref{eq:backflow_electric}), 
Eqs. (\ref{eq:spin_current_F1N_renormalized}) and (\ref{eq:spin_current_F2N_renormalized}) are replaced by 
\begin{equation}
\begin{split}
  &
  \mathbf{I}_{s}^{\rm pump}
  +
  \mathbf{I}_{s}^{\rm F_{1} \to N}
  =
  \frac{\hbar}{4\pi}
  \tilde{g}_{\rm r}^{\uparrow\downarrow}
  \mathbf{m}_{1}
  \times
  \frac{d\mathbf{m}_{1}}{dt}
\\
  &-
  \frac{1}{4\pi}
  \left[
    \tilde{g}^{*}
    \left(
      \mathbf{m}_{1}
      \cdot
      \bm{\mu}_{\rm N}
    \right)
    \mathbf{m}_{1}
    +
    \frac{h \tilde{g}^{*} I}{\tilde{g}_{e} e}
    \mathbf{m}_{1}
    +
    \tilde{g}_{\rm r}^{\uparrow\downarrow}
    \mathbf{m}_{1}
    \times
    \left(
      \bm{\mu}_{\rm N}
      \times
      \mathbf{m}_{1}
    \right)
  \right],
  \label{eq:spin_current_F1N_electric}
\end{split}
\end{equation}
\begin{equation}
\begin{split}
  \mathbf{I}_{s}^{\rm F_{2} \to N}
  =
  -\frac{1}{4\pi}
  &
  \left[
    \tilde{g}^{*}
    \left(
      \mathbf{m}_{2}
      \cdot
      \bm{\mu}_{\rm N}
    \right)
    \mathbf{m}_{2}
    -
    \frac{h \tilde{g}^{*} I}{\tilde{g}_{e} e}
    \mathbf{m}_{2}
  \right.
\\
  &\left.
  \ \ \ \ \ \ +
    \tilde{g}_{\rm r}^{\uparrow\downarrow}
    \mathbf{m}_{2}
    \times
    \left(
      \bm{\mu}_{\rm N}
      \times
      \mathbf{m}_{2}
    \right)
  \right],
  \label{eq:spin_current_F2N_electric}
\end{split}
\end{equation}
where, as done in the previous section, we assume that 
the material parameters of two ferromagnets are identical, and thus, 
omit subscripts "F${}_{k}$" from the conductances, for simplicity. 
We also assume that $g_{\rm r}^{\uparrow\downarrow} \gg g_{\rm i}^{\uparrow\downarrow}$. 
A new conductance $\tilde{g}_{e}$ is defined as 
\begin{equation}
  \frac{1}{\tilde{g}_{e}}
  =
  \frac{2 \gamma}{(1-\gamma^{2})g}
  +
  \frac{\beta}{g_{\rm sd}}
  \tanh
  \left(
    \frac{d}{2 \lambda_{\rm sd(L)}}
  \right).
\end{equation}

A characteristic current of the spin torque switching is 
the critical current of the magnetization dynamics $I_{\rm c}$, 
which can be defined as the current canceling the Gilbert damping torque 
of the F${}_{1}$ layer at the equilibrium state [\onlinecite{taniguchi08c}]. 
The equilibrium state in the present study corresponds to $\theta=0^{\circ}$. 
In this limit ($\theta \to 0$), 
Eq. (\ref{eq:alpha}) is replaced by 
\begin{equation}
  \alpha^{\prime}
  =
  \frac{\gamma_{0} \hbar \tilde{g}_{\rm r}^{\uparrow\downarrow}}{4\pi M Sd_{1}}
  \left(
    \frac{1}{2}
    -
    \frac{\pi \tilde{g}^{*}I}{e \omega \tilde{g}_{\rm r}^{\uparrow\downarrow} \tilde{g}_{e}}
  \right). 
\end{equation}
We assume that the Gilbert damping purely comes from the spin pumping. 
Then, the critical current is defined as the current satisfying $\alpha^{\prime}=0$; 
i.e., 
\begin{equation}
  I_{\rm c}
  =
  \frac{e\omega \tilde{g}_{\rm r}^{\uparrow\downarrow} \tilde{g}_{e}}{2\pi \tilde{g}^{*}}.
\end{equation}
Using the same parameter values as in the previous section, 
the critical current density $J_{\rm c}=I_{\rm c}/S$ is estimated as 
$6.3 \times 10^{6}$ A/cm${}^{2}$. 
This value is about the same order of an experimentally observed value [\onlinecite{yakushiji13}] 
($\sim 6 \times 10^{6}$ A/cm${}^{2}$ on average) 
of the critical current 
having a magnetic anisotropy field $H_{\rm K}$, 
whose magnitude (1-3 kOe) is about the same order of 
the parameter value, $\omega/\gamma_{0} \simeq 3.2$ kOe, used here. 
The dissipation due to this electric current 
based on the conventional Joule heating formula, 
$\partial Q_{A}^{\rm EC}/\partial t=\sum_{k}[\rho J_{\rm c}^{2}d_{k} + r_{{\rm F}_{k}/{\rm N}}J_{\rm c}^{2}]$, 
is evaluated as $11.8 \times 10^{3}$ fJ/(nm${}^{2}$s), 
where $r_{\rm F/N}=(h/e^{2})S/g$ 
is the F/N interface resistance. 
This value of the dissipation is two to three orders of magnitude larger than 
the dissipation due to the spin pumping 
studied in the previous section. 


We briefly investigate the origins of a large difference between the dissipations due to the spin and electric currents. 
Let us assume that the bulk and interface spin polarizations ($\beta$ and $\gamma$) are identical, 
and that the thickness of the ferromagnetic layer is much larger than the spin diffusion length ($d \gg \lambda_{\rm sd(L)}$), 
for simplicity, 
from which the critical current is simplified as 
$I_{\rm c} = e \omega \tilde{g}_{\rm r}^{\uparrow\downarrow}/(2\pi \beta)$. 
Then the ratio between the dissipations due to spin pumping and electric current becomes 
$(\partial Q_{A}^{\rm SP}/\partial t)/(\partial Q_{A}^{\rm EC}/\partial t) 
  \sim \beta^{2} \hbar/[e^{2} (\tilde{g}_{\rm r}^{\uparrow\downarrow}/S) (\rho d + r)]$. 
The square of the spin polarization, $\beta^{2}$, is on the order of $10^{-1}$. 
Also, the orders of $[(h/e^{2})S/(\tilde{g}_{\rm r}^{\uparrow\downarrow}r)]$ and $r/\rho d$ are $1$ and $0.1$, respectively. 
Then, the ratio $(\partial Q_{A}^{\rm SP}/\partial t)/(\partial Q_{A}^{\rm EC}/\partial t)$ 
is roughly $10^{-2}$, which is roughly consistent with the above evaluation. 
This consideration implies that 
a large dissipation due to the electric current 
comes from the smallness of the spin polarization. 
Also, a large bulk resistivity ($\rho$), in addition to 
the interface resistance ($r$), also contributes to 
the large dissipation due to the electric current, 
whereas only the interface resistance contributes 
to the spin pumping dissipation 
because spin pumping is an interface effect. 


To conclude this section, 
we mention that the total dissipation in the FMR consists of 
that due to spin pumping, Eq. (\ref{eq:dissipation_quantitative}), 
and that due to the intrinsic damping in the F${}_{1}$ layer. 
One can consider the possibility that 
the total dissipation in the FMR might become comparable to 
or exceed the dissipation due to the electric current (calculated above) 
when the dissipation due to intrinsic magnetic damping is included, 
despite the fact the dissipation due to spin pumping is small. 
However, we found that the intrinsic damping constant $\alpha_{0}$ should be 
at least on the order of $0.1-1$ to make the dissipation in the FMR 
comparable with that due to the electric current; see Appendix C.
On the other hand, the experimental value of the intrinsic Gilbert damping constant 
is on the order of $0.001-0.01$ [\onlinecite{oogane06}]. 
Therefore, the dissipation in the FMR is still much smaller than 
that due to the electric current 
even after the dissipation due to the intrinsic damping is taken into account. 
The energy supplied by the microwave to the F${}_{1}$ layer is 
divided into the power to sustain the magnetization precession 
and that transferred to the conduction electrons near the F${}_{1}$/N interface, 
where their ratio is roughly $\alpha_{0}:\alpha^{\prime}$. 
The former ($\propto \alpha_{0}$) is dissipated by the bulk magnetic dissipation 
whereas the latter ($\propto \alpha^{\prime}$) is dissipated by 
the spin-flip processes and spin-dependent scatterings within bulk and at the interface, 
as shown by Eq. (\ref{eq:dissipation_relation}). 


\section{Conclusion}
\label{sec:Conclusion}

The dissipation and heating due to a pure spin-current generated by spin pumping 
in a ferromagnetic / nonmagnetic / ferromagnetic multilayer 
was quantitatively investigated. 
Using spin-dependent transport theory and thermodynamics we generalized the Joule heating formula 
in the presence of spin current 
flowing in a ferromagnetic multilayer. 
The bulk and interface dissipation formulas are given 
by Eqs. (\ref{eq:dissipation_volume}) and (\ref{eq:dissipation_area}), respectively. 
For spin pumping, 
the transferred energy from the ferromagnet to the conduction electrons is not localized at the interface, 
and is dissipated throughout the system by the flow of a pure spin-current, 
as shown by Eq. (\ref{eq:dissipation_relation}). 
The dissipation due to the spin pumping, Eq. (\ref{eq:dissipation_quantitative}), 
is proportional to the enhancement of the Gilbert damping by spin pumping, Eq. (\ref{eq:alpha}). 
Using typical values of parameters in a metallic multilayer system, 
the amount of the dissipation at maximum is 
estimated to be two to three orders of magnitude smaller than 
the dissipation due to the electric current for spin torque switching. 


\section*{Acknowledgement}

The authors would like to acknowledge 
M. D. Stiles, P. M. Haney, G. Khalsa, 
R. Jansen, T. Yorozu, H. Maehara, A. Emura, T. Nozaki, 
H. Imamura, S. Tsunegi, H. Kubota, S. Yuasa, and Y. Utsumi. 
This work was supported by JSPS KAKENHI Grant-in-Aid for Young Scientists (B) 25790044. 


\appendix

\section{Non-negativity of bulk and interface dissipations}

In this Appendix, we prove that all terms 
on the right-hand side of Eq. (\ref{eq:dissipation_relation}) are positive, 
which guarantees the second law of thermodynamics; 
i.e., the dissipation, or rate of the entropy production, is positive [\onlinecite{kondepudi98}]. 
Here, we omit the subscript ``F${}_{k}$'' ($k=1,2$) from conductances, for simplicity. 


First, we prove the non-negativity of the longitudinal and transverse parts of 
the bulk dissipation. 
The longitudinal part of Eq. (\ref{eq:dissipation_volume}) can be rewritten as 
\begin{equation}
\begin{split}
  \left(
    \frac{\partial Q_{V}}{\partial t}
  \right)^{\rm L}
  &=
  \frac{J_{e}}{e}
  \frac{\partial \bar{\mu}}{\partial x}
  -
  \frac{\partial}{\partial x}
  \frac{\mathbf{J}_{s}^{\rm L}}{\hbar}
  \cdot
  \bm{\mu}^{\rm L}
\\
  &=
  -\sum_{\nu=\uparrow,\downarrow}
  j_{\nu}
  \frac{\partial \bar{\mu}_{\nu}}{\partial x}
  -
  \frac{(\bar{\mu}_{\uparrow}-\bar{\mu}_{\downarrow})}{2}
  \frac{\partial}{\partial x}
  \left(
    j_{\uparrow}
    -
    j_{\downarrow}
  \right)
\\
  &=
  \sum_{\nu=\uparrow,\downarrow}
  \frac{e^{2}}{\sigma_{\nu}}
  \left(
    j_{\nu}
  \right)^{2}
  +
  \frac{(1-\beta^{2})}{4e^{2}\rho \lambda_{\rm sd(L)}^{2}}
  \left(
    \bar{\mu}_{\uparrow}
    -
    \bar{\mu}_{\downarrow}
  \right)^{2},
\end{split}
\end{equation}
which is clearly positive. 
Here, we use the relation 
$\partial (j_{\uparrow}-j_{\downarrow})/\partial x=-(1-\beta^{2})(\bar{\mu}_{\uparrow}-\bar{\mu}_{\downarrow})/(2e^{2}\rho \lambda_{\rm sd(L)}^{2})$. 
Also, we can confirm from Eqs. (\ref{eq:diffusion_equation_transverse}) and (\ref{eq:transverse_spin_current}) that 
the transverse part, 
\begin{equation}
\begin{split}
  \left(
    \frac{\partial Q_{V}}{\partial t}
  \right)^{\rm T}
  &=
  -\frac{\partial}{\partial x}
  \frac{\mathbf{J}_{s}^{\rm T}}{\hbar}
  \cdot
  \bm{\mu}^{\rm T}
\\
  &=
    \frac{2e^{2}}{\hbar^{2} \sigma_{\uparrow\downarrow}}
  \left(
    \mathbf{J}_{s}^{\rm T}
  \right)^{2}
  +
  \frac{\sigma_{\uparrow\downarrow}}{2e^{2} \lambda_{\rm sd(T)}^{2}}
  \left(
    \bm{\mu}^{\rm T}
  \right)^{2},
  \label{eq:QVF1T}
\end{split}
\end{equation}
is positive. 
Therefore, the bulk dissipation is positive at any $x$. 


Next, let us prove the non-negativity of the interface dissipation 
by using the solutions of the spin current and spin accumulation (see also Appendix B). 
The longitudinal part of the F${}_{1}$/N interface dissipation can be written as 
\begin{equation}
  \left(
    \frac{\partial Q_{A}}{\partial t}
  \right)_{\rm F_{1}/N}^{\rm L}
  =
  \frac{\tilde{g}^{*}}{4\pi \hbar S}
  \left[
    1
    -
    \frac{\tilde{g}^{*}}{g_{\rm sd}\tanh(d_{1}/\lambda_{\rm sd(L)})}
  \right]
  \left(
    \mathbf{m}_{1}
    \cdot
    \bm{\mu}_{\rm N}
  \right)^{2}.
\end{equation}
According to Eq. (\ref{eq:g_star}), 
$1-\tilde{g}^{*}/[g_{\rm sd}\tanh(d_{1}/\lambda_{\rm sd(L)})]$ is larger than zero. 
Therefore, the longitudinal part of the F${}_{1}$/N interface dissipation is positive. 
The longitudinal part of the F${}_{2}$/N interface dissipation, 
\begin{equation}
  \left(
    \frac{\partial Q_{A}}{\partial t}
  \right)_{\rm F_{2}/N}
  =
  \frac{\mathbf{J}_{s}^{\rm F_{2} \to N}}{\hbar}
  \cdot
  \left(
    \bm{\mu}_{\rm F_{2}}
    -
    \bm{\mu}_{\rm N}
  \right),
\end{equation}
is positive because of the same reason. 
The transverse part of the F${}_{2}$/N interface dissipation, 
\begin{equation}
\begin{split}
  \left(
    \frac{\partial Q_{A}}{\partial t}
  \right)_{\rm F_{2}/N}^{\rm T}
  =
  \frac{\tilde{g}_{\rm r}^{\uparrow\downarrow}}{4\pi \hbar S}
  &
  \left\{
    1
    -
    \tilde{g}_{\rm r}^{\uparrow\downarrow}
    {\rm Re}
    \left[
      \frac{1}{g_{\rm t}\tanh(d_{2}/\ell)}
    \right]
  \right\}
\\
  &\times
  \left[
    \bm{\mu}_{\rm N}^{2}
    -
    \left(
      \mathbf{m}_{2}
      \cdot
      \bm{\mu}_{\rm N}
    \right)^{2}
  \right],
\end{split}
\end{equation}
is also positive due to similar reasons, 
where we use approximation $\tilde{g}_{\rm r}^{\uparrow\downarrow} \gg \tilde{g}_{\rm i}^{\uparrow\downarrow}$ used in Sec. \ref{sec:Evaluation of dissipation} for simplicity. 


\section{Theoretical formulas for bulk and interface dissipation}

In this Appendix, 
we discuss how to calculate the dissipation due to spin pumping 
from Eq. (\ref{eq:dissipation_relation}). 
To this end, we first show the solutions for 
the spin current and spin accumulation in the F${}_{1}$ and F${}_{2}$ layers 
because each term on the right-hand-side of Eq. (\ref{eq:dissipation_relation}) 
consists of spin current and spin accumulation, 
as shown in Eqs. (\ref{eq:dissipation_volume}) and (\ref{eq:dissipation_area}). 
The general solution for the spin current and spin accumulation are summarized in 
our previous work [\onlinecite{taniguchi08b}]. 
Here, we use these solutions, and express the spin current and spin accumulation 
in terms of the coefficients $a$ and $c$ of $\bm{\mu}_{\rm N}$ 
defined in Sec. \ref{sec:Evaluation of dissipation} 
with the assumptions $\tilde{g}_{\rm r}^{\uparrow\downarrow} \gg \tilde{g}_{\rm i}^{\uparrow\downarrow}$. 


First, we present the theoretical formulas for 
the spin current and spin accumulation within the F${}_{1}$ layer. 
We introduce two unit vectors $\mathbf{t}_{1}=\mathbf{m}_{1}\times\dot{\mathbf{m}}_{1}/|\mathbf{m}_{1} \times \dot{\mathbf{m}}_{1}|$ 
and $\mathbf{t}_{2}=-\dot{\mathbf{m}}_{1}/|\dot{\mathbf{m}}_{1}|$, 
which are orthogonal to the magnetization $\mathbf{m}_{1}$ 
and satisfy $\mathbf{t}_{1} \times \mathbf{t}_{2} = \mathbf{m}_{1}$, 
because the transverse components of the spin current and spin accumulation, 
Eqs. (\ref{eq:transverse_spin_current_def}) and (\ref{eq:transverse_spin_accumulation_def}), 
can be projected to these two directions. 
Then, the longitudinal and transverse components of the spin current in the F${}_{1}$ layer are given by 
\begin{equation}
  \mathbf{m}_{1}
  \cdot
  \mathbf{I}_{s ({\rm F}_{1})}
  =
  -\frac{\hbar \omega \tilde{g}^{*} a \sin\theta}{4\pi}
  \frac{\sinh[(x+d_{1})/\lambda_{\rm sd(L)}]}{\sinh(d_{1}/\lambda_{\rm sd(L)})},
\end{equation}
\begin{equation}
  \mathbf{t}_{1}
  \cdot
  \mathbf{I}_{s ({\rm F}_{1})}
  =
  \frac{\hbar \omega \tilde{g}_{\rm r}^{\uparrow\downarrow} (1-c) \sin\theta}{4\pi}
  {\rm Re}
  \left[
    \frac{\sinh[(x+d_{1})/\ell]}{\sinh(d_{1}/\ell)}
  \right],
\end{equation}
\begin{equation}
  \mathbf{t}_{2}
  \cdot
  \mathbf{I}_{s ({\rm F}_{1})}
  =
  \frac{\hbar \omega \tilde{g}_{\rm r}^{\uparrow\downarrow} (1-c) \sin\theta}{4\pi}
  {\rm Im}
  \left[
    \frac{\sinh[(x+d_{1})/\ell]}{\sinh(d_{1}/\ell)}
  \right].
\end{equation}
We can confirm that the 
sum of these components, 
$(\mathbf{m}_{1}\cdot\mathbf{I})\mathbf{m}_{1} + (\mathbf{t}_{1}\cdot\mathbf{I}_{s})\mathbf{t}_{1} + (\mathbf{t}_{2}\cdot\mathbf{I}_{s})\mathbf{t}_{2}$, 
at $x=0$ is identical to the spin current at the F${}_{1}$/N interface, 
$\mathbf{I}_{s}^{\rm pump}+\mathbf{I}_{s}^{\rm F_{1} \to N}$. 
Similarly, the longitudinal and transverse spin accumulation in the F${}_{1}$ layer are given by 
\begin{equation}
  \mathbf{m}_{1}
  \cdot
  \bm{\mu}_{\rm F_{1}}
  =
  \frac{\hbar \omega \tilde{g}^{*} a \sin\theta}{g_{\rm sd}}
  \frac{\cosh[(x+d_{1})/\lambda_{\rm sd(L)}]}{\sinh(d_{1}/\lambda_{\rm sd(L)})},
\end{equation}
\begin{equation}
  \mathbf{t}_{1}
  \cdot
  \bm{\mu}_{\rm F_{1}}
  =
  -\hbar 
  \omega 
  \tilde{g}_{\rm r}^{\uparrow\downarrow}
  (1-c)
  \sin\theta
  {\rm Re}
  \left[
    \frac{\cosh[(x+d_{1})/\ell]}{g_{\rm t} \sinh(d_{1}/\ell)}
  \right],
\end{equation}
\begin{equation}
  \mathbf{t}_{2}
  \cdot
  \bm{\mu}_{\rm F_{1}}
  =
  -\hbar 
  \omega 
  \tilde{g}_{\rm r}^{\uparrow\downarrow}
  (1-c)
  \sin\theta
  {\rm Im}
  \left[
    \frac{\cosh[(x+d_{1})/\ell]}{g_{\rm t} \sinh(d_{1}/\ell)}
  \right].
\end{equation}


Next, we present the explicit forms of the spin current and spin accumulation in the F${}_{2}$ layer. 
The magnetization $\mathbf{m}_{2}$ can be expressed in terms of $(\mathbf{t}_{1},\mathbf{t}_{2},\mathbf{m}_{1})$ as 
$\mathbf{m}_{2}=\cos\theta \mathbf{m}_{1} + \sin\theta \mathbf{t}_{1}$. 
We introduce two unit vectors, $\mathbf{u}_{1}=-\sin\theta\mathbf{m}_{1} + \cos\theta \mathbf{t}_{1}$ 
and $\mathbf{u}_{2}=\mathbf{t}_{2}$ 
satisfying $\mathbf{u}_{1} \times \mathbf{u}_{2} = \mathbf{m}_{2}$, 
to decompose the transverse component. 
In terms of $(\mathbf{u}_{1},\mathbf{u}_{2},\mathbf{m}_{2})$, 
$\bm{\mu}_{\rm N}$ can be expressed as 
$\bm{\mu}_{\rm N}=\hbar \omega \sin\theta [(a \cos\theta + c \sin\theta) \mathbf{m}_{2} + (-a \sin\theta + c \cos\theta) \mathbf{u}_{1}]$. 
Then, the longitudinal and transverse spin currents are given by 
\begin{equation}
\begin{split}
  \mathbf{m}_{2}
  \cdot
  \mathbf{I}_{s ({\rm F}_{2})}
  =&
  -\frac{\hbar \omega \tilde{g}^{*} (a \sin\theta \cos\theta + c \sin^{2}\theta)}{4\pi}
\\
  &\times
  \frac{\sinh[(x-d_{2})/\lambda_{\rm sd(L)}]}{\sinh(d_{2}/\lambda_{\rm sd(L)})}, 
\end{split}
\end{equation}
\begin{equation}
\begin{split}
  \mathbf{u}_{1}
  \cdot
  \mathbf{I}_{s ({\rm F}_{2})}
  =&
  -\frac{\hbar \omega \tilde{g}_{\rm r}^{\uparrow\downarrow} (-a \sin^{2}\theta + c \sin\theta \cos\theta)}{4\pi}
\\
  &\times
  {\rm Re}
  \left[
    \frac{\sinh[(x-d_{2})/\ell]}{\sinh(d_{2}/\ell)}
  \right],
\end{split}
\end{equation}
\begin{equation}
\begin{split}
  \mathbf{u}_{2}
  \cdot
  \mathbf{I}_{s ({\rm F}_{2})}
  =&
  -\frac{\hbar \omega \tilde{g}_{\rm r}^{\uparrow\downarrow} (-a \sin^{2}\theta + c \sin\theta \cos\theta)}{4\pi}
\\
  &\times
  {\rm Im}
  \left[
    \frac{\sinh[(x-d_{2})/\ell]}{\sinh(d_{2}/\ell)}
  \right].
\end{split}
\end{equation}
We can confirm that the 
sum of these components, 
$(\mathbf{m}_{2}\cdot\mathbf{I})\mathbf{m}_{2} + (\mathbf{u}_{1}\cdot\mathbf{I}_{s})\mathbf{u}_{1} + (\mathbf{u}_{2}\cdot\mathbf{I}_{s})\mathbf{u}_{2}$, 
at $x=0$ is identical to the spin current at the F${}_{2}$/N interface, 
$-\mathbf{I}_{s}^{\rm F_{2} \to N}$. 
The longitudinal and transverse spin accumulations are given by 
\begin{equation}
\begin{split}
  \mathbf{m}_{2}
  \cdot
  \bm{\mu}_{\rm F_{2}}
  =&
  \frac{\hbar \omega \tilde{g}^{*} (a \sin\theta \cos\theta + c \sin^{2}\theta)}{g_{\rm sd}}
\\
  &
  \times
  \frac{\cosh[(x-d_{2})/\lambda_{\rm sd(L)}]}{\sinh(d_{2}/\lambda_{\rm sd(L)})},
\end{split}
\end{equation}
\begin{equation}
\begin{split}
  \mathbf{u}_{1}
  \cdot
  \bm{\mu}_{\rm F_{2}}
  =&
  \hbar 
  \omega 
  \tilde{g}_{\rm r}^{\uparrow\downarrow}
  (-a \sin^{2}\theta + c \sin\theta \cos\theta)
\\
  &\times
  {\rm Re}
  \left[
    \frac{\cosh[(x-d_{2})/\ell]}{g_{\rm t} \sinh(d_{2}/\ell)}
  \right],
\end{split}
\end{equation}
\begin{equation}
\begin{split}
  \mathbf{u}_{2}
  \cdot
  \bm{\mu}_{\rm F_{2}}
  =&
  \hbar 
  \omega 
  \tilde{g}_{\rm r}^{\uparrow\downarrow}
  (-a \sin^{2}\theta + c \sin\theta \cos\theta)
\\
  &\times
  {\rm Im}
  \left[
    \frac{\cosh[(x-d_{2})/\ell]}{g_{\rm t} \sinh(d_{2}/\ell)}
  \right].
\end{split}
\end{equation}



\begin{figure}
\centerline{\includegraphics[width=0.5\columnwidth]{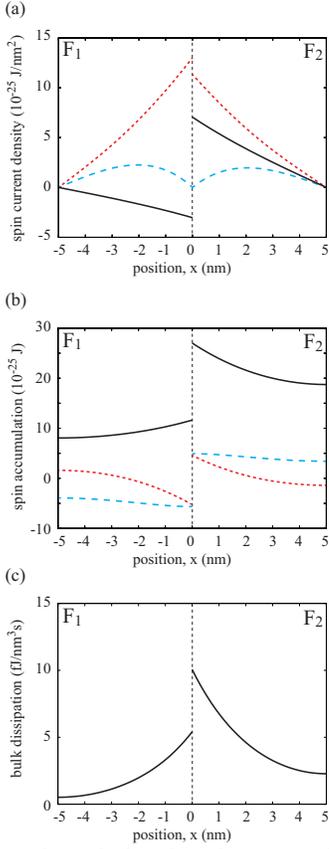}}\vspace{-3.0ex}
\caption{
         Examples of the distributions of 
         (a) longitudinal (solid) and transverse (dotted and dashed) spin current densities, 
         (b) longitudinal (solid) and transverse (dotted and dashed) spin accumulations, 
         and (c) bulk dissipations for $\theta=45^{\circ}$. 
         \vspace{-3ex}}
\label{fig:fig4}
\end{figure}



Figures \ref{fig:fig4} (a) and (b) show 
the spatial distributions of the spin current density and spin accumulation, respectively. 
The spin current density and spin accumulation are decomposed into the longitudinal and transverse directions, 
where the solid lines correspond to the longitudinal components 
whereas the dotted ($\parallel \mathbf{t}_{1}$ or $\mathbf{u}_{1}$) and dashed ($\parallel \mathbf{t}_{2}$ or $\mathbf{u}_{2}$) 
correspond to the transverse components. 
The values of the parameters are identical to those used in Sec. \ref{sec:Evaluation of dissipation} 
with $\theta=45^{\circ}$. 
Because spin pumping occurs at the F${}_{1}$/N interface, 
the spin current density and spin accumulation are concentrated near this interface. 
We emphasize that the spatial directions of the longitudinal and transverse spin are 
different between the F${}_{1}$ and F${}_{2}$ layers 
when the magnetizations, $\mathbf{m}_{1}$ and $\mathbf{m}_{2}$, are noncollinear; 
as a result the spin current in Fig. \ref{fig:fig4} (a) looks discontinuous at the interface, 
although Eq. (\ref{eq:conservation_spin_current}) is satisfied. 


We now consider the dissipation formulas. 
The longitudinal and transverse parts of the bulk dissipation in the F${}_{1}$ layer can be expressed as 
\begin{equation}
\begin{split}
  \left(
    \frac{\partial Q_{V}}{\partial t}
  \right)_{\rm F_{1}}^{\rm L}
  =&
  \frac{\hbar \omega^{2}}{4\pi S}
  \frac{\tilde{g}^{* 2} a^{2} \sin^{2}\theta}{g_{\rm sd} \lambda_{\rm sd(L)} \sinh^{2}(d_{1}/\lambda_{\rm sd(L)})}
\\
  & \times 
  \cosh
  \left[
    \frac{2(x+d_{1})}{\lambda_{\rm sd(L)}}
  \right], 
  \label{eq:QVF1L_eq}
\end{split}
\end{equation}
\begin{equation}
\begin{split}
  \left(
    \frac{\partial Q_{V}}{\partial t}
  \right)_{\rm F_{1}}^{\rm T}
  =&
  \frac{\hbar \omega^{2} \tilde{g}_{\rm r}^{\uparrow\downarrow 2}(1-c)^{2} \sin^{2}\theta}{4\pi S^{2}}
  \frac{e^{2}}{h \sigma_{\uparrow\downarrow}}
\\
  &\times
  \left\{
    \frac{1}{\lambda_{\rm sd(T)}^{2}}
    \bigg|
      \frac{\ell \cosh[(x+d_{1})/\ell]}{\sinh(d_{1}/\ell)}
    \bigg|^{2}
  \right.
\\
  &
  \left.
    \ \ \ \ +
    \bigg|
      \frac{\sinh[(x+d_{1})/\ell]}{\sinh(d_{1}/\ell)}
    \bigg|^{2}
  \right\}.
  \label{eq:QVF1T_eq}
\end{split}
\end{equation}
Similarly, the longitudinal and transverse parts of the bulk dissipation in the F${}_{2}$ layer 
can be expressed as 
\begin{equation}
\begin{split}
  \left(
    \frac{\partial Q_{V}}{\partial t}
  \right)_{\rm F_{2}}^{\rm L}
  =&
  \frac{\hbar \omega^{2}}{4\pi S}
  \frac{\tilde{g}^{* 2} (a \sin\theta \cos\theta + c \sin^{2}\theta)^{2}}{g_{\rm sd} \lambda_{\rm sd(L)} \sinh^{2}(d_{2}/\lambda_{\rm sd(L)})}
\\
  &\times
  \cosh
  \left[
    \frac{2(x-d_{2})}{\lambda_{\rm sd(L)}}
  \right]. 
  \label{eq:QVF2L_eq}
\end{split}
\end{equation}
\begin{equation}
\begin{split}
  \left(
    \frac{\partial Q_{V}}{\partial t}
  \right)_{\rm F_{2}}^{\rm T}
  =&
  \frac{\hbar \omega^{2} \tilde{g}_{\rm r}^{\uparrow\downarrow 2} (-a \sin^{2}\theta + c \sin\theta \cos\theta)^{2}}{4\pi S^{2}}
  \frac{e^{2}}{h \sigma_{\uparrow\downarrow}}
\\
  &\times
  \left\{
    \frac{1}{\lambda_{\rm sd(T)}^{2}}
    \bigg|
      \frac{\ell \cosh[(x-d_{2})/\ell]}{\sinh(d_{2}/\ell)}
    \bigg|^{2}
  \right.
\\
  &
  \left.
  \ \ \ \ +
    \bigg|
      \frac{\sinh[(x-d_{2})/\ell]}{\sinh(d_{2}/\ell)}
    \bigg|^{2}
  \right\}.
  \label{eq:QVF2T_eq}
\end{split}
\end{equation}
Figure \ref{fig:fig4} (c) shows the spatial distribution of the bulk dissipation, 
which is also concentrated near the interface. 


The longitudinal part of the F${}_{1}$/N interface dissipation 
and the longitudinal and transverse parts of the F${}_{2}$/N interface dissipations 
are given by 
\begin{equation}
  \left(
    \frac{\partial Q_{A}}{\partial t}
  \right)_{\rm F_{1}/N}^{\rm L}
  =
  \frac{\hbar \omega^{2} \tilde{g}^{*} a^{2} \sin^{2}\theta}{4\pi S}
  \left[
    1
    -
    \frac{\tilde{g}^{*}}{g_{\rm sd}\tanh(d_{1}/\lambda_{\rm sd(L)})}
  \right], 
\end{equation}
\begin{equation}
\begin{split}
  \left(
    \frac{\partial Q_{A}}{\partial t}
  \right)_{\rm F_{2}/N}^{\rm L}
  =&
  \frac{\hbar \omega^{2} \tilde{g}^{*} (a \sin\theta \cos\theta + c \sin^{2}\theta)^{2}}{4\pi S}
\\
  & \times
  \left[
    1
    -
    \frac{\tilde{g}^{*}}{g_{\rm sd}\tanh(d_{2}/\lambda_{\rm sd(L)})}
  \right], 
\end{split}
\end{equation}
\begin{equation}
\begin{split}
  \left(
    \frac{\partial Q_{A}}{\partial t}
  \right)_{\rm F_{2}/N}^{\rm T}
  =&
  \frac{\hbar \omega^{2} \tilde{g}_{\rm r}^{\uparrow\downarrow} (-a \sin^{2}\theta + c \sin\theta \cos\theta)^{2}}{4\pi S}
\\
  &\times
  \left\{
    1
    -
    \tilde{g}_{\rm r}^{\uparrow\downarrow}
    {\rm Re}
    \left[
      \frac{1}{g_{\rm t}\tanh(d_{2}/\ell)}
    \right]
  \right\}.
\end{split}
\end{equation}


For $\theta=45^{\circ}$, we quantitatively evaluate that 
$\int_{-d_{1}}^{0} dx (\partial Q_{V}/\partial t)_{\rm F_{1}}^{\rm L}=3.34$ fJ/(nm${}^{2}$s), 
$\int_{-d_{1}}^{0} dx (\partial Q_{V}/\partial t)_{\rm F_{1}}^{\rm T}=6.51$ fJ/(nm${}^{2}$s), 
$\int_{0}^{d_{2}} dx (\partial Q_{V}/\partial t)_{\rm F_{2}}^{\rm L}=18.15$ fJ/(nm${}^{2}$s), 
and 
$\int_{0}^{d_{2}} dx (\partial Q_{V}/\partial t)_{\rm F_{2}}^{\rm T}=4.95$ fJ/(nm${}^{2}$s), 
respectively. 
Also, the interface dissipations are 
quantitatively evaluated as $(\partial Q_{A}/\partial t)_{\rm F_{1}/N}^{\rm L}=0.44$ fJ/(nm${}^{2}$s), 
$(\partial Q_{A}/\partial t)_{\rm F_{2}/N}^{\rm L}=2.39$ fJ/(nm${}^{2}$s), 
and $(\partial Q_{A}/\partial t)_{\rm F_{2}/N}^{\rm T}=8.03$ fJ/(nm${}^{2}$s) for $\theta=45^{\circ}$, respectively. 
We can confirm that the value of the dissipation evaluated from these values as Eq. (\ref{eq:dissipation_relation}) 
is the same with that evaluated from Eq. (\ref{eq:dissipation_FNF}) with Fig. \ref{fig:fig3}. 


\section{Dissipation due to intrinsic damping}

In this Appendix, we briefly evaluate the dissipation due to 
the magnetization precession in the FMR experiment, 
which arises from the intrinsic Gilbert damping. 
In the FMR, the energy supplied by the microwave balances with the dissipation due to the damping, 
and the magnetization precesses practically on the constant energy curve. 
The magnetization dynamics with the macrospin assumption is described by 
the Landau-Lifshitz-Gilbert (LLG) equation 
\begin{equation}
  \frac{d \mathbf{m}_{1}}{dt}
  =
  -\gamma_{0}
  \mathbf{m}_{1}
  \times
  \mathbf{H}
  -
  \alpha_{0}
  \gamma_{0}
  \mathbf{m}_{1}
  \times
  \left(
    \mathbf{m}_{1}
    \times
    \mathbf{H}
  \right),
  \label{eq:LLG}
\end{equation}
where the magnetic field $\mathbf{H}$ relates to the magnetic energy density $E$ via $\mathbf{H}=-\partial E/\partial (M \mathbf{m}_{1})$. 
From Eq. (\ref{eq:LLG}), 
the change of the energy density averaged on the constant energy curve is given by 
\begin{equation}
\begin{split}
  \overline{
    \frac{dE}{dt}
  }
  &\equiv
  \frac{1}{\tau}
  \oint dt 
  \frac{dE}{dt}
\\
  &=
  -\frac{\alpha \gamma_{0} M}{\tau}
  \oint dt 
  \left[
    \mathbf{H}^{2}
    -
    \left(
      \mathbf{m}_{1}
      \cdot
      \mathbf{H}
    \right)^{2}
  \right],
  \label{eq:dEdt}
\end{split}
\end{equation}
where, $\tau=\oint dt$ is the precession period on a constant energy curve. 
Assuming that the ferromagnet has uniaxial anisotropy $\mathbf{H}=(0,0,H_{\rm K}m_{z})$ as done in Sec. \ref{sec:Comparison with spin torque switching}, 
Eq. (\ref{eq:dEdt}) is given by 
\begin{equation}
  \overline{\frac{dE}{dt}}
  =
  -\alpha_{0}
  \gamma_{0}
  M H_{\rm K}^{2}
  \sin^{2}\theta
  \cos^{2}\theta.
\end{equation}
The microwave should supply the energy density $-\overline{dE/dt}$ to sustain the precession. 
Then, the energy supplied by the microwave per unit area per unit time is $\alpha_{0}\gamma_{0}MH_{\rm K}^{2}d_{1}\sin^{2}\theta\cos^{2}\theta$, 
where $d_{1}$ is the thickness of the ferromagnet. 
Comparing this energy with the dissipation due to the spin pumping carried by the spin current, Eq. (\ref{eq:dissipation_quantitative}), 
the ratio of the dissipation between the intrinsic damping and spin pumping is 
\begin{equation}
  \frac{| \overline{dE/dt}|d_{1}}{\partial Q_{A}^{\rm SP}/\partial t}
  \sim
  \frac{\alpha_{0}}{\alpha^{\prime}},
\end{equation}
where $\alpha^{\prime}$ is given by Eq. (\ref{eq:alpha}). 
The dissipation due to the spin pumping ($\propto \alpha^{\prime}$) is two to three orders of magnitude 
smaller than the dissipation due to the electric current. 
Therefore, the intrinsic Gilbert damping constant $\alpha_{0}$ giving bulk magnetic dissipation 
of the same order of magnitude as the dissipation due to the electric current 
is roughly $10^{2-3} \times \alpha^{\prime}$. 
From the value of $\alpha^{\prime}$ in Fig. \ref{fig:fig3} (b), 
this gives an $\alpha_{0}$ on the order of $0.1-1$. 




\begin{thebibliography}{47}
\bibitem[{com()}]{comment1}
\bibinfo{note}{The definition of the spin accumulation obeys Refs.
  \cite{tserkovnyak02b}. A cautionary note: Some papers define
  $\bm{\mu}/2$ as the spin accumulation. According to this latter definition,
  the longitudinal spin accumulation is $\bm{\mu}^{\rm
  L}=[(\bar{\mu}_{\uparrow}-\bar{\mu}_{\downarrow})/2]\mathbf{m}$, and
  similarly, the amount of the transverse spin accumulation is a half of that
  by our definition. Accordingly, $\hbar$ in Eqs. (\ref{eq:heat_current}),
  (\ref{eq:dissipation_volume}), and (\ref{eq:dissipation_area}) should be
  replaced by $\hbar/2$.}
\end{thebibliography}


\end{document}